\documentclass[
aps,
pre,
twocolumn,
amsmath,
amssymb,
showpacs,
floatfix,
letterpaper
]{revtex4-1}
\usepackage[english]{babel}
\usepackage{graphicx}
\usepackage{bm}
\usepackage{amssymb,amsmath}
\usepackage{dcolumn}

\begin{document}

\title{Correlation between rearrangements and soft modes in polymer
  glasses during deformation and recovery}

\author{Anton \surname{Smessaert}} \author{J\"{o}rg \surname{Rottler}}
\email{jrottler@physics.ubc.ca}
\affiliation{Department of Physics and Astronomy, The University of
  British Columbia, 6224 Agricultural Road, Vancouver, British
  Columbia, Canada, V6T 1Z1}

\begin{abstract}
We explore the link between soft vibrational modes and local
relaxation events in polymer glasses during physical aging, active
deformation at constant strain rate, and subsequent recovery. A
softness field is constructed out of the superposition of the
amplitudes of the lowest energy normal modes, and found to predict up
to 70\% of the rearrangements. Overlap between softness and
rearrangements increases logarithmically during aging and recovery
phases as energy barriers rise due to physical aging, while yielding
rapidly rejuvenates the overlap to that of a freshly prepared glass.
In the strain hardening regime, correlations rise for uniaxial tensile
deformation but not for simple shear. These trends can be explained by
considering the differing degrees of localization of the soft modes in
the two deformation protocols.
\end{abstract}
\pacs{64.70.P, 81.05.Kf, 61.43.Fs, 61.43.Bn}
\date{\today}
\maketitle

\section{Introduction}
The mechanical behavior of solids is often successfully described by
continuum mechanics of a homogeneous medium.  Polymer glasses,
however, exhibit structural heterogeneity at the nanoscale. For
instance, local elastic shear moduli evaluated in small volume
elements containing several tens of monomers can differ substantially
from the bulk macroscopic value \cite{yoshimoto_mechanical_2004}. As a
result, amorphous materials can be separated into stiff and soft
regions. At the same time, glassy dynamics at the monomer level is
heterogeneous, with groups of particles relaxing rapidly in a
collective manner while other regions remain essentially immobile
\cite{aichele_polymer-specific_2003,berthier_dynamical_2011}. Understanding
the existence or absence of a link between local structure and
dynamics is one of the important open questions in the physics of
disordered materials.

Recent molecular simulations of a variety of model glass formers have
provided strong evidence that the locations of structural
rearrangements correlate with measures of structural
heterogeneity. For instance, simulatons of a model polymer glass under
tensile deformation show that nonaffine residual monomer displacements
are largest in regions with small positive shear moduli
\cite{papakonstantopoulos_molecular_2008}. Simulation studies of the
athermal deformation of amorphous two-dimensional binary mixtures also
indicate that the nonaffine displacement field resulting from the
macroscopic shear deformations of a few percent is directly
related to the spatial structure of the elastic moduli
\cite{tsamados_local_2009,mizuno_measuring_2013}, while little
correlation is found to other structural variables such as local
density.

An alternative description of structural heterogeneity that also
untilizes a harmonic description has been developed around the notion
of soft vibrational modes. Amorphous solids generally exhibit an
excess amount of low energy modes (Boson peak), and a large number of
them are quasilocalized, i.e. they involve only very few particles
\cite{laird_localized_1991}. The particles involved in these modes
exhibit structural differences in their neighborhood configuration
\cite{schober_localized_1991}, suggesting that they might play the
role of defects in crystals. Recently, it has become possible to
identify the structural signatures of rearranging particles with
generalized structure functions that describe radial and bond
orientations \cite{cubuk_identifying_2015}. A ``softness field''
formed by the superposition of the amplitudes of the lowest frequency
normal modes also yields a heterogeneous partitioning into hard and
soft regions.  Multiple recent simulation studies in supercooled
liquids
\cite{widmer-cooper_irreversible_2008,widmer-cooper_localized_2009,hocky_small_2013}
and amorphous packings under shear at zero
\cite{tanguy_vibrational_2010,manning_vibrational_2011,mosayebi_soft_2014}
or finite temperature \cite{schoenholz_understanding_2014} confirm
that regions of large vibrational amplitude overlap with the loci of
structural rearrangements. An advantage of the soft mode description
over the elastic moduli approach is that the soft modes also
anticipate the direction of particle motion in a rearrangement
\cite{smessaert_structural_2014,rottler_predicting_2014}. This
correlation is robust and insensitive to the specific model system or
diagnostic of the rearrangement. Low energy sound waves therefore
scatter off flow defects and provide information about those
collective particle motions that are most easily excited.

In a recent contribution, we explored quantitatively the robustness of
this softness field for the prediction of monomer relaxation in aging
polymer glasses \cite{smessaert_structural_2014} for different
temperatures and aging times. We showed that rearrangements identified
as rapid changes in the particle positions (hops out of local cages)
are up to seven times more likely at the softest regions, where the
direction of motion also aligns near perfectly with the local
polarization vectors.  These correlations are only erased once more
than 50\% of the entire system has undergone rearrangements.

In the present work, we report a study of the correlation between
irreversible rearrangements and local softness in actively deformed
polymer glasses. This is important because previous studies have
focused only on steadily sheared two-dimensional monomeric glasses
formers
\cite{tanguy_vibrational_2010,manning_vibrational_2011,schoenholz_understanding_2014},
while polymers are distinctly three-dimensional and exhibit strain
hardening. Here we subject the polymer to uniaxial tensile deformation
(extensional flow) and compare to the simple shear protocol used in
previous studies. Moreover, we explore subsequently correlations
between soft regions and local relaxation events in the regimes of
aging, deformation, and recovery. During the aging phase, correlations
grow due to a descent into deeper energy minima, but return quickly to
their as-quenched values once plastic flow sets in. In the uniaxial
tensile deformation protocol, correlations rise again with increasing
strain, which we show to arise from an increase in the degree of
localization of the soft modes as the aspect ratio of the sample
changes. In simple shear by contrast, this effect is absent and
correlations remain near constant during deformation. During the
recovery phase, correlations increase only weakly as inherent
structure energies return to their undeformed values.

\section{Methods}\label{methods}
We study the polymer glass using molecular dynamics techniques and the
well-known finitely extensible nonlinear elastic (FENE) bead-spring
model~\cite{kremer_dynamics_1990}, which is an excellent computer
glass-former~\cite{bennemann_molecular-dynamics_1998,smessaert_recovery_2012}.
Each linear polymer consists of 50 identical monomers with covalent
bonds along the polymer backbone that were modeled by a non-linear
stiff spring-like interaction. Inter- and intra-chain interactions
between not covalently bonded monomers are modeled with a 6-12
Lennard-Jones (LJ) potential, which was truncated and
force-shifted~\cite{allen_computer_1989} with a cut-off of 2.5 times
the bead diameter for computational efficiency, and to ensure that the
Hessian is defined. The latter is important for the analysis of the
vibrational spectrum. All quantities are stated in the usual LJ-units
based on well energy $u$, particle diameter $a$, mass $m$ and
characteristic time scale $\tau_{LJ}=\sqrt{ma^2/u}$, which is just
below twice the mean collision time. The simulation time step was set
to $\Delta t=0.0075\tau_{LJ}$, and all results are averaged over 20
independent simulation runs.

We simulate N=10,000 particles comprising 200 polymers in an initially
cubic simulation box with periodic boundary conditions. Our simulation
protocol is a direct extension of our earlier study that quantified
the correlation between soft modes and irreversible particle
rearrangements called hops in quiescent polymer
glasses~\cite{smessaert_structural_2014}. The glass is generated by
rapidly quenching an equilibrated melt with density $\rho=1.043$ at
constant volume (NVT) from $T=1.2$ to $T=0.3$ with a constant quench
rate of $\dot{T}=6.7\cdot 10^{-4}$. For the remainder of the
simulation, the temperature is held constant. In a first step, the
glass is aged without deformation at zero pressure (NPT) to the age
$t_{age}=7,500,000$ at $T=0.3$, which is just below the glass
transition temperature. In the uniaxial tensile deformation protocol,
the glass is then deformed in one direction with a constant strain
rate $\dot{\epsilon}=10^{-5}$, while the pressure perpendicular to the
deformation axis is kept at zero. As a result, the simulation box
shape changes from cubic to rectangular. The deformation ends at a
final engineering strain $\epsilon(t)=[L_z(t)-L_z(0)]/L_z(0)=4$ with
$L_z$ being the simulation box size along the deformation axis. In a
final step, both deformation and barostat were turned off so that the
system recovers at fixed volume and the stresses relax. This
post-deformation recovery regime was explored for a time
$t_{r}=150,000$. The simple shear protocol imposes a constant shear
rate of the same magnitude by deforming the simulation box into a
parallelepiped. Here no additional barostating is performed.

Two key measurements are performed during the simulation: At
different times during aging, deformation and recovery, we first
store a configuration snapshot for the identification of the soft
modes and then detect irreversible particle rearrangements in the
whole simulation box immediately following the time of the
snapshot. As introduced in full detail in
Ref.~\cite{smessaert_structural_2014}, we calculate a softness
measure for each particle based on its participation in the low energy
vibrational modes. This calculation was done in multiple steps:
Starting from a snapshot, we first found the inherent structure using
a combination of gradient descent and damped dynamics
(FIRE~\cite{bitzek_structural_2006}) algorithms with a minimal total
force criterion. Then, the Hessian
\begin{equation}\label{eq:hessian}
H_{(\mathbf{r}_{i})_{k}(\mathbf{r}_{j})_{l}} = \frac{\partial^{2}U(\{\mathbf{r}_i\})}{\partial (\mathbf{r}_{i})_{k}\partial (\mathbf{r}_{j})_{l}}\;.
\end{equation}
is calculated from the inherent structure particle locations
$\mathbf{r}$ as well as the potential energy $U$, and the $N_m=600$ lowest
energy eigenmodes are calculated using ARPACK. The softness of a particle is
defined as the superposition of the participation fractions in the low
energy vibrational
modes~\cite{widmer-cooper_irreversible_2008,widmer-cooper_localized_2009,mosayebi_soft_2014},
\begin{equation}\label{eq:softnessfield}
\phi_{i}=\frac{1}{N_{m}}\sum_{j=1}^{N_{m}}|\mathbf{e}_{j}^{(i)}|^{2}\;.
\end{equation}
Here, the polarization vector $\mathbf{e}_{j}^{(i)}$ is the projection
of the eigenvector of mode $j$ on the degrees of freedom of particle
$i$. The softness field $\phi$ depends on a single parameter, the
number of low energy modes $N_m$, and the scaling factor is
added to make the softness an intensive quantity in terms of $N_{m}$.
An optimal value of $N_m=600$ (2\% of the modes) is chosen based on
our previous analysis \cite{smessaert_structural_2014}. A particle $i$
is "softer" the larger $\phi_{i}$, which is used to rank the particles
according to their relative softness. The absolute value of $\phi$ is
not in itself meaningful, since the participation fractions are
normalized quantities, i.e., $\sum_{i=1}^N|\mathbf{e}_j^{(i)}|^{2}=1$.

\begin{figure}[tb]
	\includegraphics{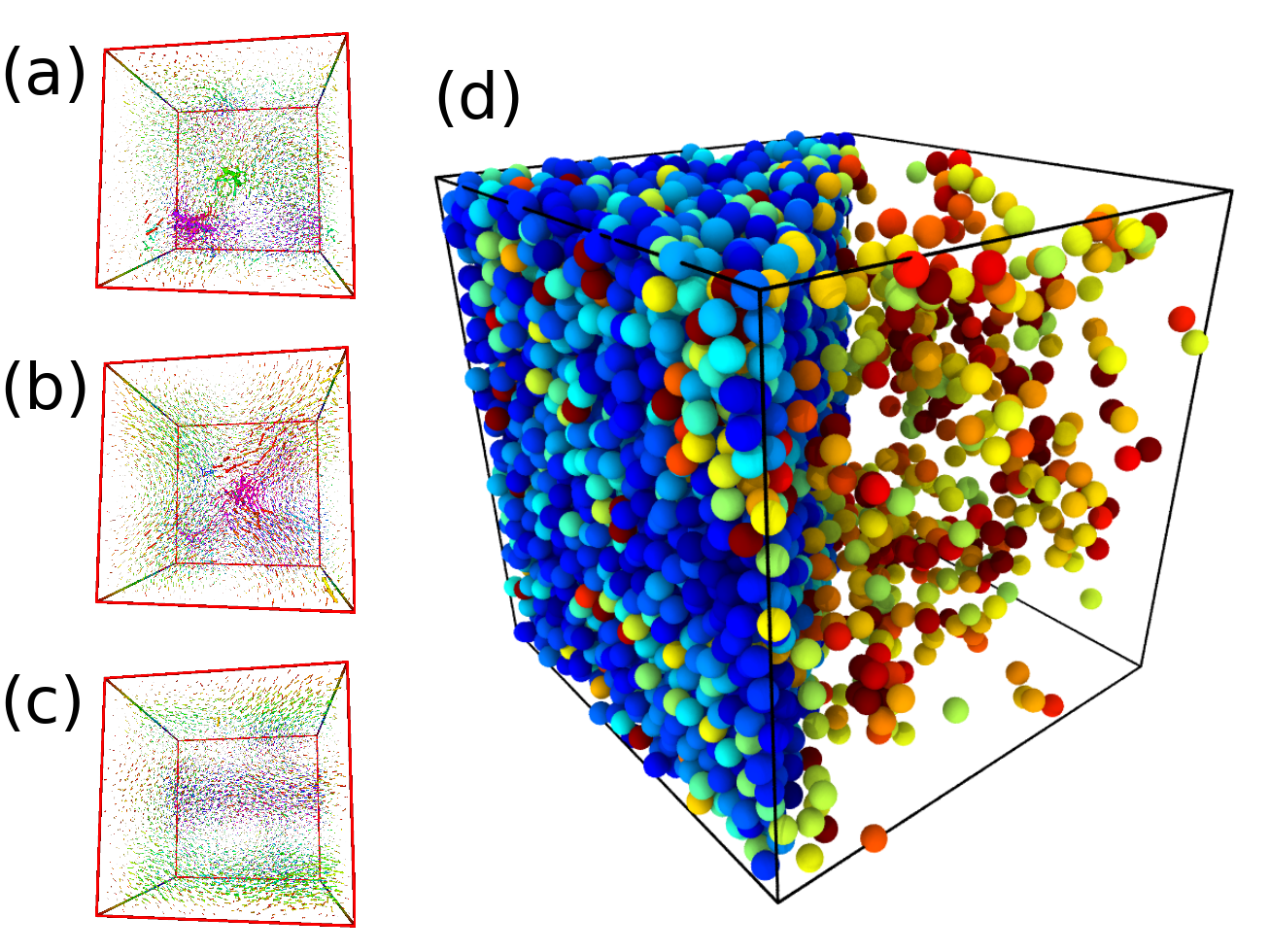}
	\caption{\label{fig:eigenmodes}(a-c) Three exemplary lowest
          energy vibrational modes, visualized by their polarization
          vector fields (coloring indicates depth). The participation
          ratios as defined in Eq.~\eqref{eq:participationRatio} are
          0.12(a), 0.29(b) and 0.62(c). (d) Exemplary simulation
          snapshot with particle softness indicated by color (blue to
          red - small to large $\phi$). The right side shows only the
          10\% softest particles.}
\end{figure}

Figure~\ref{fig:eigenmodes}(a-c) visualizes three exemplary low energy
eigenmodes calculated at the end of the aging period by plotting their
polarization vector fields. One can see that the modes (a) and (b) are
quasi-localized, since large polarization vectors are concentrated on
a small number of particles that are spatially clustered. The spectrum
also exhibits extended, planar-wave like modes and an example is shown
in panel (c). Panel (d) shows a snapshot of the system where each
particle is colored according to its softness (blue is small $\phi$
and red is large $\phi$). On the right side of the simulation box, we
only show the 10\% softest particles, illustrating the heterogeneous
spatial distribution.
\begin{figure*}[tb]
\centering
\includegraphics{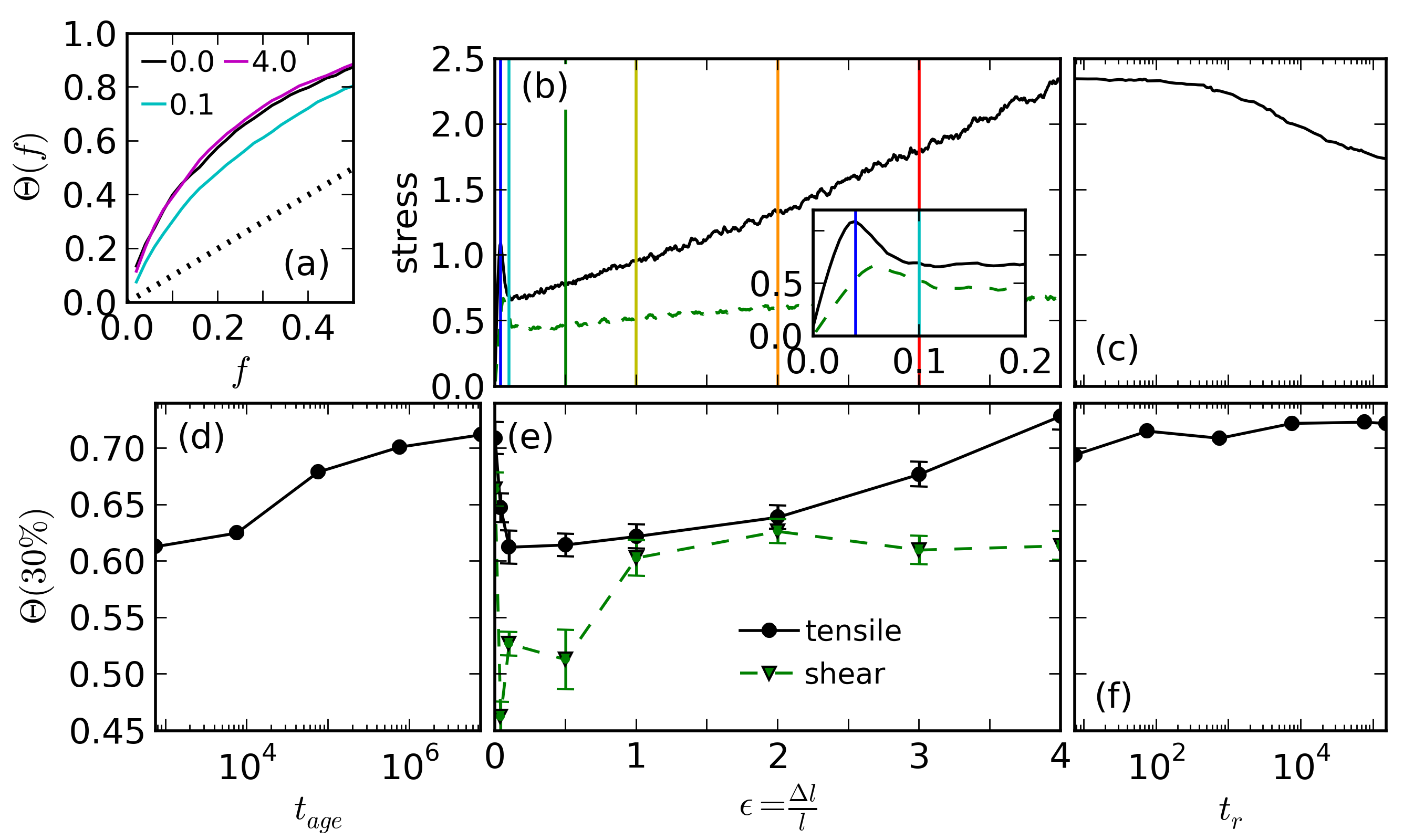}
\caption{\label{fig:correlation_deformation}(a) Fraction of hops in
  soft spots as function of coverage fraction of soft spots measured
  in three deformation regimes for uniaxial tension: elastic
  ($\epsilon=0.0$), strain softening ($\epsilon=0.1$), and strain
  hardening ($\epsilon=4.0$). The dotted line indicates no
  correlation. (b) Stress-strain curves for uniaxial tensile (solid
  line) and simple shear (dashed line). Vertical colored lines
  indicate the investigated deformations at strains
  $\epsilon=0.0,\,0.04,\,0.1,\,0.5,\,1.0,\,2.0,\,3.0,\,4.0$ and the
  inset shows the region of the yield point in more detail. (c) Stress
  recovery from uniaxial tensile deformation. The evolution of the
  predictive success rate reached at $f=0.3$ is shown during aging
  (d), deformation (e) and in the recovery regime (f).}
\end{figure*}

Irreversible particle rearrangements are identified using a
detection algorithm that was introduced in a previous
study~\cite{smessaert_distribution_2013}, where it is explained in
greater detail. In short, we monitor the trajectory of each particle
on-the-fly during the simulation and isolate rapid changes or
\emph{hops} of particles by calculating the mean distance squared
\begin{equation}
P_{hop}^{(i)}=\sqrt{\left\langle\left(\mathbf{r}^{A}_i - \bar{r}^{B}_{i}\right)^2\right\rangle_{A}\cdot \left\langle\left(\mathbf{r}^{B}_i - \bar{r}^{A}_{i}\right)^2\right\rangle_{B}}
\end{equation}
between the earlier (A) and later (B) half of the trajectory of
particle $i$ in a time window that moves with the simulation. Here,
the averages $\langle . \rangle_{A}$ [$\langle . \rangle_{B}$] are
taken over all trajectory points in A [B] and $\bar{r}^{A}$
[$\bar{r}^{B}$] is the mean position in the respective trajectory
segment. A hop is detected when $P_{hop}>P_{th}=0.21$, and we record
particle identifier, time of hop, as well as initial and final
position. The value of the threshold is related to the plateau in the
mean square displacement \cite{smessaert_distribution_2013}. To
isolate the irreversible rearrangements we exclude back-and-forth hops
of a particle between the same two positions. This is done by removing
a sequence of two hops of the same particle, if the final position of
the second hop is within a distance of $\sqrt{P_{th}}/2$ of the
initial position of the first hop.

\section{Results}
We discuss the spatial correlation between softness field and monomer
hops in terms of a predictive success rate $\Theta$
\cite{smessaert_structural_2014}. Here, the softness field is
binarized into a soft spot map denoted $\phi_i^{(b)}$ by assigning a
softness of one to the fraction $f$ of particles with largest softness
and zero to the other particles. We then calculate the fraction of the
first $N_h=100$ hopping particles that are part of a soft spot, or
\begin{equation*}
\Theta(f) = \frac{\sum_{i=1}^N \phi_i^{(b)} h_i}{N_h}
\end{equation*}
with $h_i=1$ if particle $i$ is one of the first $N_h$ particles to
hop after the measurement of the softness field, and $h_i=0$
otherwise.

Figure~\ref{fig:correlation_deformation}(a) shows the predictive
success rate measured at three different strains during the uniaxial
tensile deformation (qualitatively similar curves are found under
simple shear). Stress-strain curves for the two deformation modes are
shown in Fig.~\ref{fig:correlation_deformation}(b) and reveal
post-yield strain hardening in both cases, albeit to a much lower
degree in the simple shear case. It can be seen that the three strains
shown in panel (a) correspond to the onset of deformation at the end
of the aging period, right after the yield peak, and the end of the
strain hardening regime.  The dotted line indicates $\Theta$ for
randomly distributed soft spots (no correlation) for comparison. For
all strains, we find a larger than random correlation between softness
field and the occurrence of hops. 

To facilitate the analysis of trends in the spatial correlation during
aging, deformation and recovery, we now focus on the predictive
success rate at a coverage fraction of 30\%.  This is the fraction
where the difference between measured $\Theta$ and uncorrelated value
(dashed line) is maximal. In Fig.~\ref{fig:correlation_deformation}(d)
we show the evolution of $\Theta(30\%)$ during the initial aging
period.  In agreement with the results reported in
Ref.~\cite{smessaert_structural_2014}, the correlation increases from
$0.61$ to $0.71$ as the age grows by four orders of magnitude. Panel
(e) shows the predictive success rate measured at different points
during uniaxial tensile and simple shear deformation. In the tensile
case (solid line), the spatial correlation in the elastic regime at
$\epsilon=0$ is roughly equal to the value found in the quiescent
state immediately prior to the deformation. At the yield strain
$\epsilon=0.04$, the correlation has decreased to $0.64$ and at the end
of the strain softening regime ($\epsilon=0.1$) $\Theta(30\%)$ has
reached the pre-aging value $0.61$.  Even larger decreases in fact
below the pre-aging value are seen in the pure shear case. This
reversal of the aging effects is consistent with mechanical
rejuvenation, by which deformation strains exceeding the yield strain
erase the thermal history and return the material to a freshly quenched
glassy state \cite{smessaert_recovery_2012}.

Interestingly, trends in spatial correlation begin to diverge between
the two deformation modes in the strain hardening regime.  Under
uniaxial tension, we find that $\Theta(30\%)$ monotonically increases
with growing strain. The increase accelerates at large strains
$\epsilon>2$, reaching a value of $0.72$ at $\epsilon=4$, which is
above the predictive success rate measured in the quiescent state
prior to the deformation. By contrast, the correlation during simple
shear saturates at the value of the freshly quenched glass for strains
larger than one. In the recovery regime, investigated only for the
tensile case (panel (f)), $\Theta(30\%)$ first drops slightly but then
increases again weakly as the material continues to age.

\begin{figure}[t]
\centering
\includegraphics{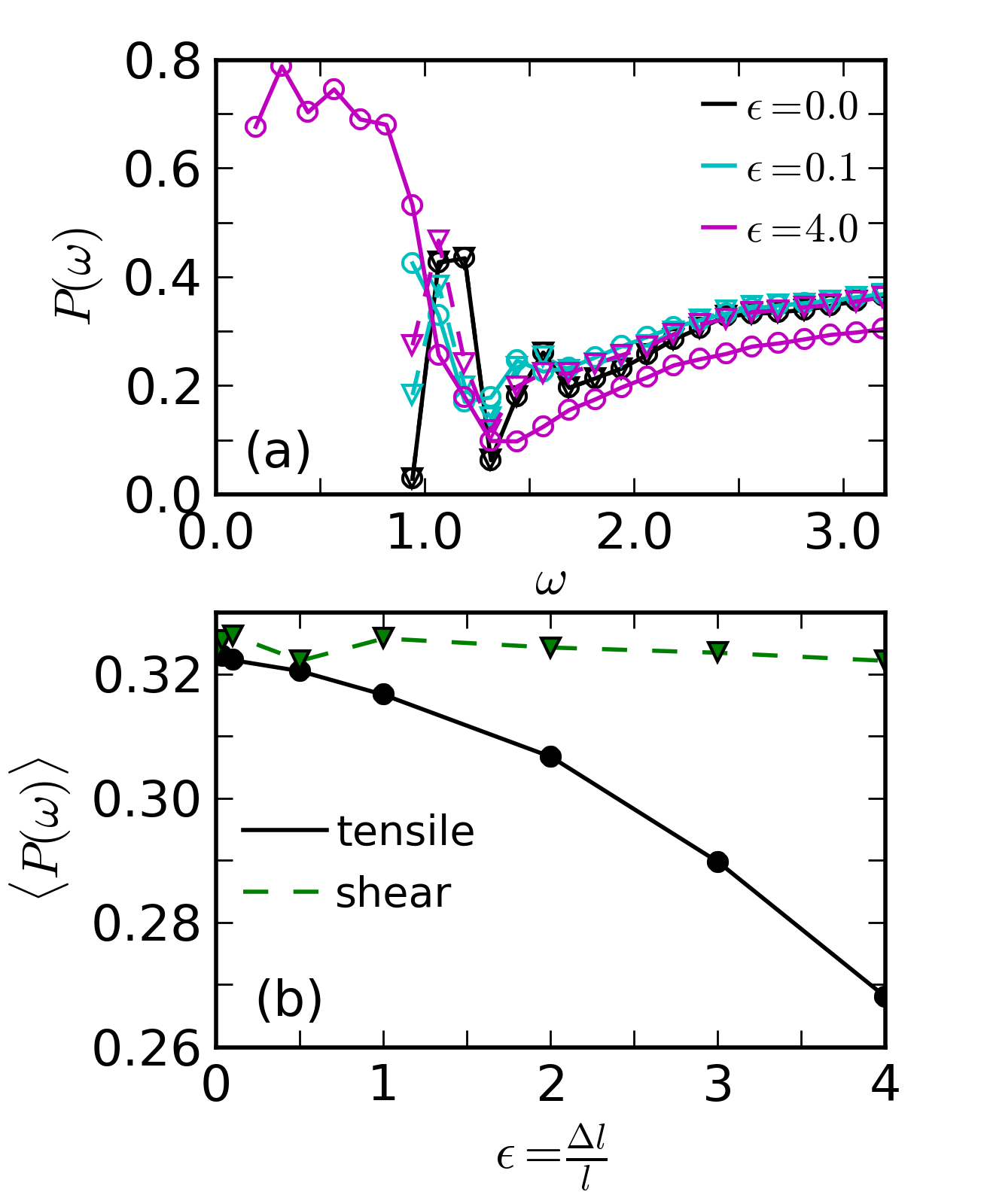}
\caption{\label{fig:participationratio}(a) Mean participation ratio as
  function of eigenfrequency at three values of strain  for uniaxial tensile
  $(\circ)$ and simple shear $(\triangledown)$ deformations..  (b) Mean
  participation ration of all $N_m$ modes used for the softness field
  calculation as function of total strain.}
\end{figure}

\begin{figure*}[t]
\centering
\includegraphics{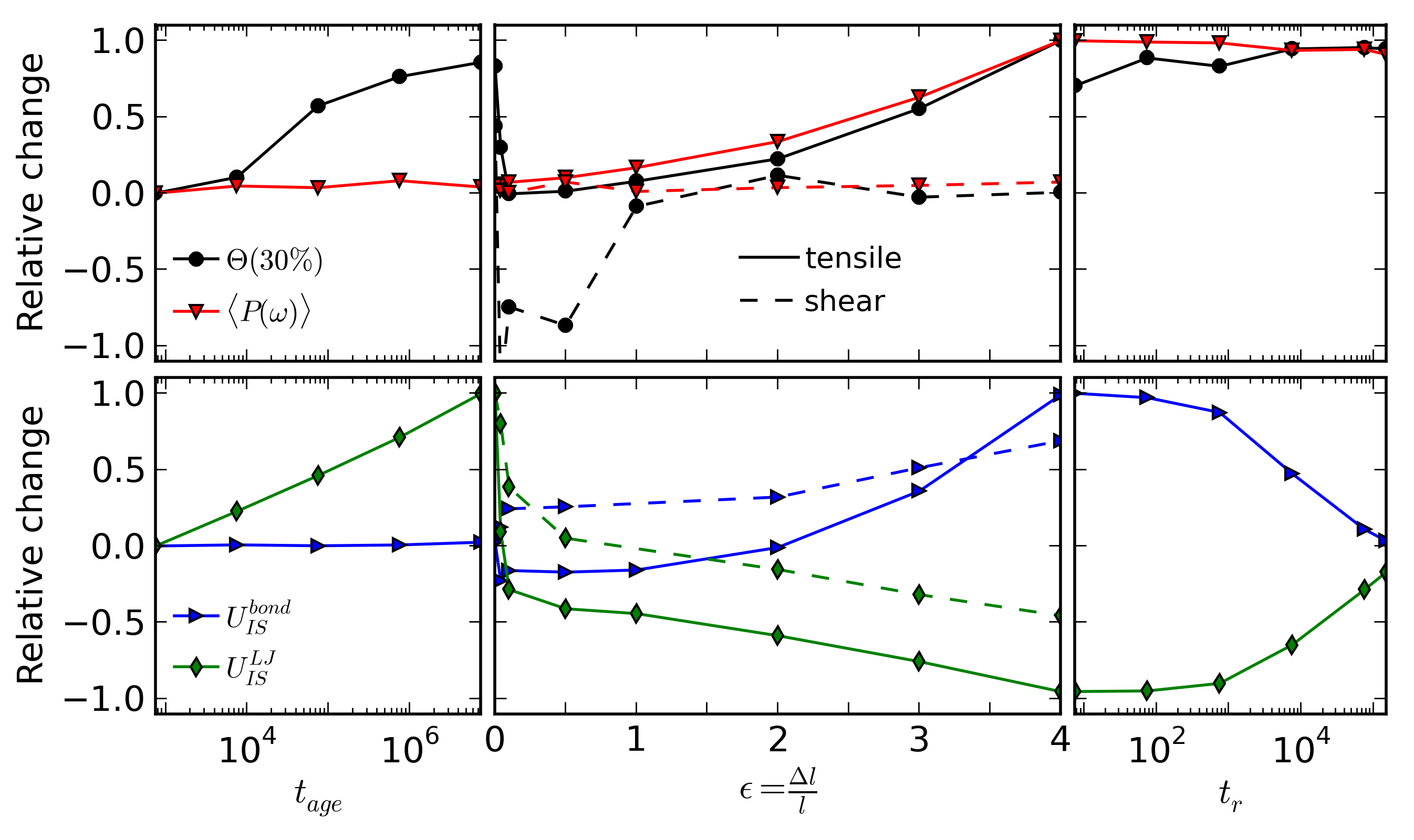}
\caption{\label{fig:relativechange}Comparison of spatial correlation,
  participation ratio (top row) and potential energy stored in the LJ
  interaction and in the covalent bonds (bottom row) during aging,
  deformation and recovery. The relative change is defined as
  $(Q-Q_0)/(Q_{\rm ext}-Q_0)$, where Q is a placeholder for one of the
  measured quantities (see legend), and $Q_{\rm ext}$ is an extremal
  value observed during the simulation run (maximum for
  $\Theta(30\%)$, $U_{IS}^{\rm bond}$ and minimum for $\langle
  P(\omega)\rangle$, $U_{IS}^{LJ}$).}
\end{figure*}

What is the origin of the different behavior of the
softness-rearrangement correlation in tensile vs.~simple shear
deformation? In order to answer this question, we analyze the low
energy vibrational mode spectrum during the deformation period more
carefully. The extent of localization of a mode $j$ can be quantified
by the participation ratio
\begin{equation}\label{eq:participationRatio}
P_j=\frac{\left(\sum_{i=1}^{N}(\mathbf{e}_{j}^{(i)})^{2}\right)^{2}}{N\sum_{i=1}^{N}(\mathbf{e}_{j}^{(i)})^{4}}\;.
\end{equation}
A value of $P_{j}=1$ means that all particles are participating
equally in mode $j$, whereas a small value indicates that the mode is
quasi-localized around a few active particles. In
Fig.~\ref{fig:participationratio} (a) we show the participation ratio
$P(\omega)$ as function of mode frequency $\omega$. Since the
vibrational spectrum is a feature of the inherent structure,
$P(\omega)$ at strain $\epsilon=0$ reflects the quiescent state at the
end of the aging period. The participation ratios at the end of the
strain softening regime ($\epsilon=0.1$) are nearly unchanged, with a
slight shift of the extended modes (large participation ratio) towards
smaller frequencies for both deformation modes. 

Important differences emerge in the strain hardening regime at
$\epsilon=4$.  While $P(\omega)$ for simple shear is unchanged,
uniaxial tension deformation results in two new features: First, we
find modes with large participation ratio at much smaller
frequencies. This change is due to the large (400\%) elongation of the
simulation box along the deformation axis in our uniaxial tension
simulation, which allows modes with larger wavelength to "fit" into
the simulation volume. In general one expects extended modes to emerge
on the low energy side of the spectrum as the system dimension
increases, since the material looks increasingly homogeneous on large
length scales. At same time, however, we find that the participation
ratios of the modes at $\omega > 1.5$ are \emph{reduced} compared to
the undeformed system. With increasing linear dimension, the
likelihood of finding a low energy quasilocalized mode is therefore
enhanced. This effect is absent in the simple shear case, during which
the simulation box periods do not change.

In fig.~\ref{fig:participationratio}(b) we show the evolution of the
mean participation ratio $\langle P(\omega)\rangle$ of all $N_m=600$
modes used for the calculation of the softness field with
deformation. For tensile deformation, we find a gradual reduction of
the average participation ratio, indicating that the soft modes become
more and more localized as the deformation grows. By contrast, the
average participation ratio does not change during simple shear. These
results suggest that the change in participation ratio is merely a
consequence of the changing box dimensions during uniaxial strain. In
order to separate deformation effects from geometry, we have
equilibrated the same polymer glass in a simulation box that has
identical shape to a deformed polymer at 400\% strain and computed the
distribution of participation ratio at zero load. These distributions
are identical to those found at the end of the hardening regime, which
confirms that the change in $\langle P(\omega)\rangle$ is of
geometrical origin.

We now show that the increasing degree of localization of the soft
modes under uniaxial deformation explains the rise of the overlap
$\Theta(30\%)$ between softness and rearrangements. To this end, we
compare the evolution of $\Theta(30\%)$ to changes in the mean
participation ratio in the top panels of
Fig.~\ref{fig:relativechange}.  To facilitate the comparison, we
introduce a scaling that normalizes these quantities by the extremal
change during the simulation run. The relative change of a quantity Q
with regard to a reference value $Q_0$ from the freshly quenched state
is defined as $(Q-Q_0)/(Q_{\rm ext}-Q_0)$, where $Q_{\rm ext}$ is the
extremal value measured during the aging, deformation and in the
recovery regime. For the spatial correlation we use the maximal value,
while for the mean participation ratio we use the minimal value.  The
top left panel shows that the mean participation ratio does not change
during aging, while the overlap $\Theta(30\%)$ rises roughly
logarithmically.  The overlap between hops and soft modes improves as
the system reaches deeper energy wells and the energy barriers for
rearrangement rise with age. During uniaxial tensile deformation
(middle panel), we find that after the initial rejuvenation drop,
$\Theta(30\%)$ rises in lockstep with the mean participation ratio
$\langle P(\omega)\rangle$. This observation suggests that the
improving overlap of $\Theta(30\%)$ in the hardening regime is indeed
due to an increased amount of highly localised modes. During simple
shear by contrast, neither overlap nor mean participation ratio
increase with strain. $\langle P(\omega)\rangle$ is again constant in
the recovery regime where the simulation box size does not change.

The bottom panels of Fig.~\ref{fig:relativechange} provide as
additional information the evolution of the inherent structure energy,
separated into contributions from pairwise LJ interactions and
covalent bonds. During aging (left panel), we see the well-known
logarithmic aging increase of $U_{IS}^{LJ}$ while the mean bond energy
$U_{IS}^{\rm bond}$ remains constant. These trends mirror those in the
panel above for the overlap.  In the deformation stage (middle panel),
$U_{IS}^{LJ}$ decreases as the glass is pulled up higher on the energy
landscape, while $U_{IS}^{\rm bond}$ increases as more and more energy
is stored in the covalent bonds of the aligning polymer chains.  While
this increase of $U_{IS}^{\rm bond}$ with strain is also broadly
consistent with the rise of the overlap measure in the uniaxial
tensile deformation, it disagrees in simple shear deformation where
the overlap is constant. Moreover, in the recovery regime (right
panel), $U_{IS}^{\rm bond}$ drops rapidly while $\Theta(30\%)$ is very
weakly increasing. These observations provide further evidence that
strain hardening does not influence the softness map - rearrangement
correlations. Our recovery regime is just long enough for both
inherent structure energies to reach their undeformed values.

\section{Discussion}
We have investigated the correlation between local relaxation events,
identified as irreversible monomer hops out of local cages, and a
softness map constructed from the superposition of low energy
vibrational modes, in polymer glasses during physical aging,
deformation and structural recovery. In the as-quenched glass, hops
are about twice as likely to occur on soft spots than elsewhere in the
polymer. This correlation increases logarithmically to about 2.4 over
4 decades of aging time and rises because physical aging amplifies the
influence of the potential energy landscape on the dynamics. Both
simple shear and uniaxial tensile deformation rapidly rejuvenate this
correlation once plastic flow sets in. Further uniaxial tensile
deformation increases the soft spot-relaxation overlap again due to an
increase in quasilocalized modes as the sample elongates in the
tensile direction, while no such effect occurs in simple shear. Strain
hardening, although present in our model polymer, does not seem to
play a major role in these trends. Accordingly, the present effects
can be expected in short molecule glass formers as well.

The present results indicate that two factors control the ability of
soft modes to predict structural rearrangement: position on the energy
landscape, and degree of localization of the lowest energy modes. The
lower the inherent structure energies and the higher the barriers, the
better the ability of the potential energy landscape (a T=0 property)
to predict dynamics at elevated temperature near the glass transition.
Additionally, the extent of overlap also depends on system
size. Although the longest wavelength modes tend to have plane wave
character, increasing the system size simultaneously generates more
highly localized modes below the Boson peak, and these are most
efficient in finding soft spots. These results call for a systematic
study of finite size effects on the low energy vibrational spectrum of
disordered solids and their relationship to rearrangements.

\section*{Acknowledgments}
We thank E.~del Gado for helpful discussions. This work was supported
by the Natural Sciences and Engineering Research Council of Canada
(NSERC). Computing time was provided by Compute Canada.

\end{document}